\documentclass{aastex}
\usepackage{graphics}
\usepackage{amssymb}
\usepackage{emulateapj5}
\usepackage{apjfonts}

\makeatletter
\newenvironment{tablehere}
  {\def\@captype{table}}
  {}

\newenvironment{figurehere}
  {\def\@captype{figure}}
  {}
\makeatother

\begin{document}

\submitted{Accepted in ApJL. to be appear in Dec. 10th issue}
\title{Lithium Depletion Boundary in a Pre-Main Sequence Binary System}

\author{Inseok Song\altaffilmark{1,2}, M. S. Bessell\altaffilmark{3}, B.
Zuckerman\altaffilmark{2}}

\altaffiltext{1}{UCLA Center for Astrobiology Research Scientist}

\altaffiltext{2}{Dept. of Physics and Astronomy --- University of
California, Los Angeles\\
Los Angeles, CA 90095--1562, USA}

\altaffiltext{3}{Research School of Astronomy and Astrophysics\\
Institute of Advanced Studies\\
The Australian National University, ACT 2611, Australia}

\begin{abstract}
A lithium depletion boundary is detected in HIP~112312 (GJ~871.1 A and
B), a \( \sim 12 \)~Myr old pre-main sequence binary system.  A strong
(EW\( \approx 300 \)~m\AA{}) Li~6708~\AA{} absorption feature is seen
at the secondary (\( \sim \)M4.5) while no Li~6708~\AA{} feature is
detected from the primary (\( \sim \)M4). The physical companionship
of the two stars is confirmed from common proper motions. Current
theoretical pre-main sequence evolutionary models cannot
simultaneously match the observed colors, brightnesses, and Li
depletion patterns of this binary system. At the age upper limit of
20~Myr, contemporary theoretical evolutionary models predict too slow
Li depletion. If true Li depletion is a faster process than predicted
by theoretical models, ages of open clusters (Pleiades, \( \alpha
\)~Persei, and IC~2391) estimated from the Li depletion boundary
method are all overestimated.  Because of the importance of the open
cluster age scale, development of self-consistent theoretical models
to match the HIP~112312 data is desirable.
\end{abstract}


\section{Introduction}

Gravitationally contracting low mass (\( M\lesssim 1\, M_{\odot } \))
stars are fully convective (Hayashi track evolution) until they near
the main sequence. As the contraction proceeds, the stellar
temperature rises and when the core temperature reaches \( \sim
2.5\times 10^{6} \)~K, depletion of lithium begins via \(
^{7}Li(p,\alpha )^{4}He \) reaction \citep[][ and references
therein]{Ventura}. Since the lithium burning process is very sensitive
to the core temperature (burning rate~\( \propto T^{20} \),
\citealt{Bildsten}), once Li burning starts, it will consume all
available lithium in a very short timescale (\( \lesssim 10 \)~Myr).
However, more massive stars (\( M\gtrsim 1\, M_{\odot } \)) form
radiative cores in their early evolution that results in a separation
of the surface layers from the hotter inner regions. As long as the
temperature at the base of the convective layer is below the Li
burning temperature, surface Li can be preserved. On the other hand,
if stellar mass is too low (\( \lesssim 0.065\, M_{\odot } \)), a star
(in fact, a brown dwarf) can never attain a high enough core
temperature to burn Li; hence such brown dwarfs should always show
strong Li absorption at 6708~\AA{}. For the higher mass brown dwarfs
(mass range of \( 0.065\, M_{\odot }-0.072\, M_{\odot } \)), lithium
eventually can burn \citep{BasriARAA}. 

Because of the fast Li burning process, the location of the Li burning
boundary in a coeval stellar group should be well defined. The Li
depletion boundary (LDB) of older clusters stays at
$0.065\,M_{\odot}$, since that is the mass below which brown dwarfs
never get hot enough to burn lithium. This implies that the LDB age
determination only works up to an age of about 250~Myr which is the
age when the interior of a $\sim0.065\,M_{\odot}$ brown dwarf just
creeps above the lithium burning temperature.  LDBs of younger
clusters will be located at higher masses than the LDB of old clusters
because lesser mass stars take more time to reach the lithium burning
temperature by gravitational contraction than more massive stars do.
Once the luminosity of cluster stars at the LDB is known, the LDB
location can be compared to that of theoretical models from which we
can estimate the cluster age.  To date, LDBs have been detected in
three open clusters --- Pleiades \citep{Basri,Stauffer98}, \( \alpha
\)~Persei \citep{Stauffer99}, and IC~2391 \citep{Barrado99}.
Interestingly, ages determined from the LDB method for these clusters
are all \( \sim 50 \)\% larger than traditional upper main sequence
fitting ages (Table~\ref{LDB}), and the lithium ages have generally
been adopted as correct.  If so, then traditional interior models for
the upper main sequence must be revised.

\begin{table*}
\caption{Li depletion boundaries in open clusters\label{LDB}}
{\centering \begin{tabular}{lcc|cccl}
\hline 
\multicolumn{1}{c}{cluster}&
\multicolumn{2}{c}{age (Myr)}&
\multicolumn{3}{c}{Li depletion boundary}&
\multicolumn{1}{c}{Reference}\\
\cline{2-3} \cline{4-6} 
\multicolumn{1}{c}{name}&
UMS\( ^{a} \)&
LDB&
sp.&
M\( _{I_{c}} \)&
\( (R-I)_{c} \)&
\\
\hline
IC~2391&
35&
53&
M5.0&
10.25&
1.91&
\citet{Barrado99}\\
\( \alpha  \) Persei&
70&
90&
M6.0&
11.47&
2.12&
\citet{Stauffer99}\\
Pleiades\( ^{\dag } \)&
85&
125&
M6.5&
12.20&
2.20&
\citet{Stauffer98}\\
\hline
\multicolumn{7}{l}{\( ^{a} \) These  are upper main sequence fitting ages}\\
\multicolumn{7}{l}{\( ^{\dag } \) LDB of the Pleiades was first discovered by \citet{Basri}
at slightly earlier spectral type}\\
\end{tabular}\par}
\end{table*}

We have detected the LDB in a pre-main sequence (PMS) binary system
HIP~112312 from medium and high resolution spectroscopic observations.
To date, this is the only system other than open clusters whose LDB is
detected. The HIP~112312 binary system has a well determined distance
from Hipparcos (\( d=24\, pc \)) and \( BVI_{c}JHK_{s} \) photometric
magnitudes. Using these data, we can check if theoretical models of
photometric isochrone ages and Li depletion ages are self-consistent.
In this paper, we compare frequently used theoretical models with
parameters derived from our observations.

\section{Data}

\subsection{Observed Data}

\begin{figure*}
\begin{minipage}[t]{\columnwidth}
{\centering \resizebox*{0.95\columnwidth}{!}{\includegraphics{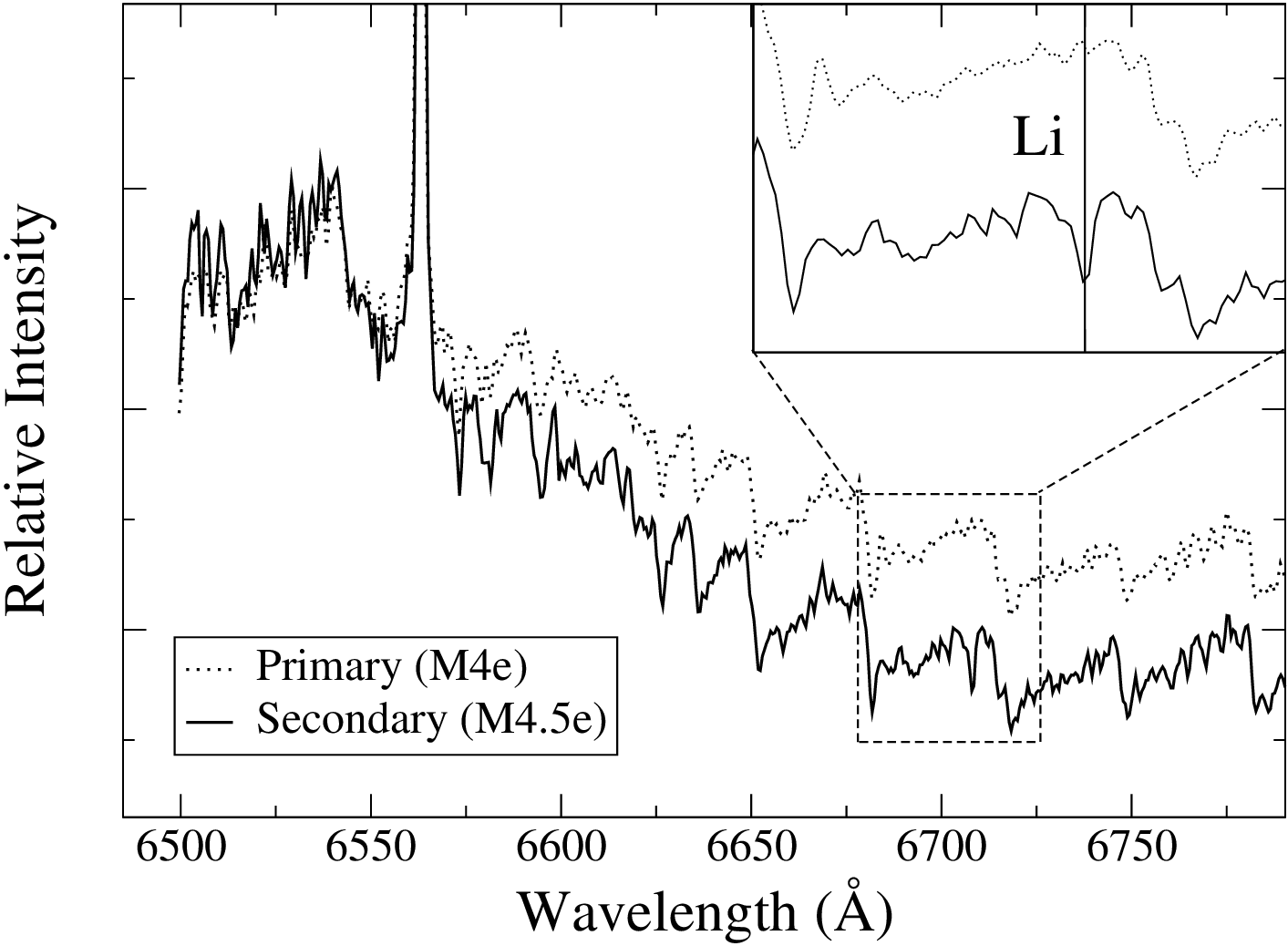}} \par}
\caption{Part of the DBS spectra of the HIP~112312 system. The dotted line
represents the primary spectrum and the solid line is for the secondary.
The secondary spectrum shows steeper slope than the primary spectrum
not because of the temperature difference, but presumably due to
stronger molecular bands (i.e., CaH). \label{spectrum}}
\end{minipage}
\hfill
\begin{minipage}[t]{\columnwidth}
{\centering \resizebox*{0.95\columnwidth}{!}{\includegraphics{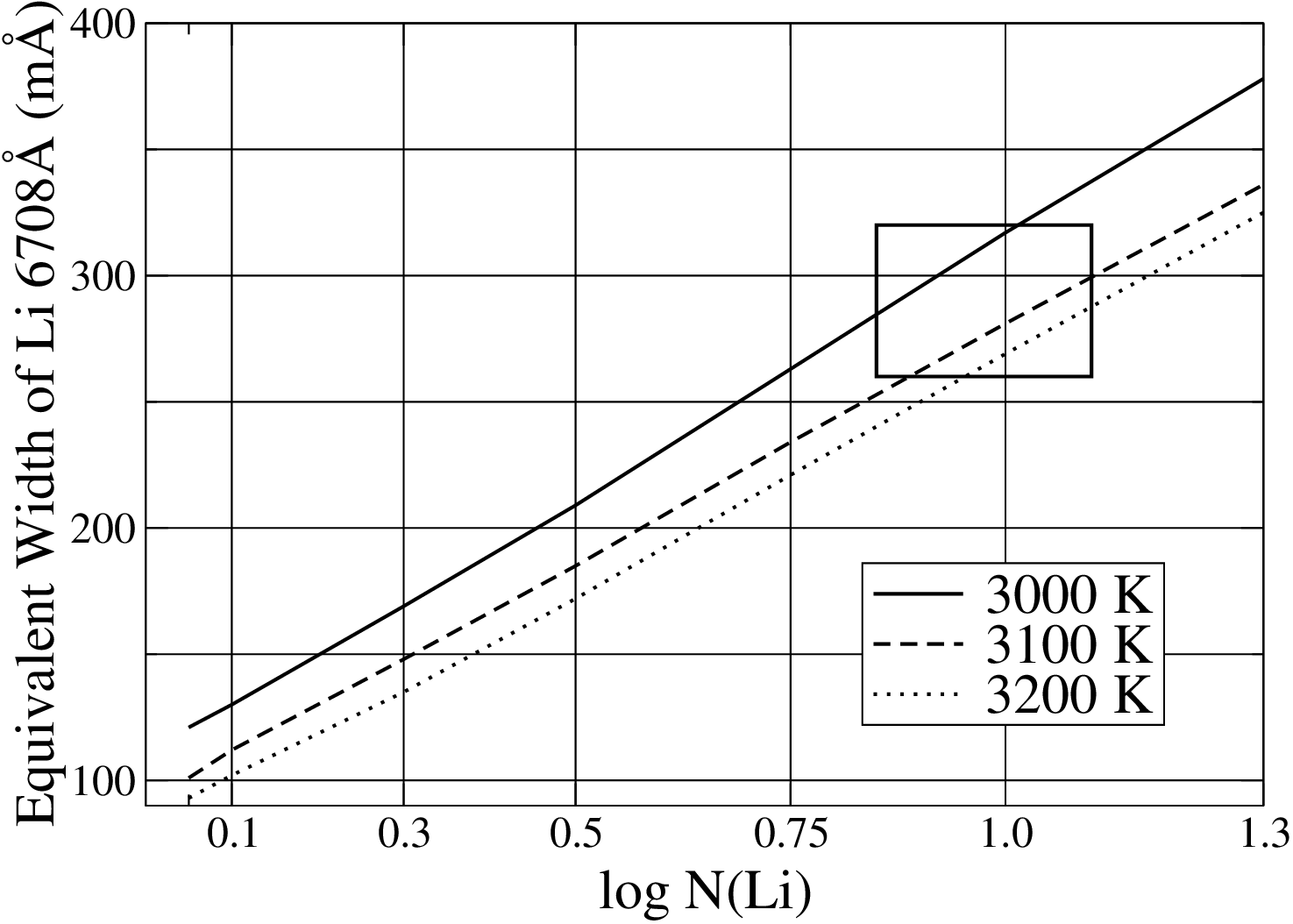}} \par}
\caption{Li~6708\,\AA{} curves of growth for $\log g=4.3$ constructed
from a set of synthetic model spectra (\citealt{PHEONIX} and see text
for details). The rectangular box represents an acceptable range of Li
abundance for the secondary star of HIP~112312.  \label{LiCurve}}
\end{minipage}
\end{figure*}

We obtained spectra of WW~PsA (\( = \)GJ~871.1~A) and TX~PsA (\( =
\)GJ~871.1~B) as part of an extensive on-going search for young and
nearby stars to Earth (Song, Bessell, \& Zuckerman, ApJ in prep.).
High resolution spectra were obtained using an echelle spectrograph
on the Nasmyth-B focus and medium resolution spectra were obtained
with Double Beam Spectrograph (DBS) on the Nasmyth-A focus of the
Australian National University's 2.3~m telescope. The red channel of
the DBS covered the spectral range $6500-7450$\,\AA\ at a measured
resolution of 1.2\,\AA{} (0.55\,\AA{}/pixel).  DBS spectra,
displayed in Figure~\ref{spectrum}, have $\sim5000$ counts per pixel
in the vicinity of 6700\,\AA{}. From these spectra, one can clearly
see the non-detection and detection of the Li~6708\,\AA\ absorption
feature from the primary and secondary, respectively, and strong
H$\alpha$ emission from both stars.  Eight orders of the echelle
covered portions of the spectra between 5800 and 7230\,\AA{}. At
orders containing the H$\alpha$ and Li 6708\,\AA\ lines, the
measured resolution was 0.45\,\AA{} (0.17\,\AA{}/pixel). 

All spectra were reduced following standard procedure using IRAF. We
also obtained \( BVI_{c} \) photometric magnitudes of the binary as
part of a larger photometry program (Shobbrook et al., in preparation)
for young stars identified in our on-going spectroscopy program. The
HIP~112312 components have \( JHK_{s} \) magnitudes from the 2MASS 2nd
release database. Spectroscopic and photometric data are summarized in
Table~\ref{data}.

\subsection{Derived Data}

\subsubsection{\protect\( T_{eff}\protect \) and spectral types \label{Teff}}

Some colors of the HIP~112312 binary from \( BVI_{c}JHK_{s} \) are
slightly inconsistent compared to normal M dwarfs. For example, the \(
B-V \) of the two stars are bluer by about 0.1~mag than normal M
dwarfs.  The lower gravity of a young M dwarf compared to the mean
gravities of ZAMS M dwarfs could account for much of this difference
and a few stars in the Taurus-Auriga region studied by \citet{KH95}
have similarly blue \( B-V \) colors for their \( V-I \) colors.  
The Pleiades M dwarfs are also bluer in $B-V$ than one would expect
for their $V-I$ or $V-K$ colors (Stauffer et al. 2002, in prep.). This is probably
\begin{tablehere}
\caption{Characteristics of HIP~112312 binary system\label{data}}
{\centering \begin{tabular}{cr @{$\pm$}lr @{$\pm$}l}
\hline 
&
\multicolumn{2}{c}{GJ 871.1A}&
\multicolumn{2}{c}{GJ 871.1 B}\\
&
\multicolumn{2}{c}{WW PsA}&
\multicolumn{2}{c}{TX PsA}\\
\hline
RA (2000)&
\multicolumn{2}{c}{\( 22:44:57.94 \)}&
\multicolumn{2}{c}{\( 22:45:00.04 \)}\\
DEC (2000)&
\multicolumn{2}{c}{\( -33:15:01.7 \)}&
\multicolumn{2}{c}{\( -33:15:26.0 \)}\\
Sp. type&
\multicolumn{2}{c}{M4e}&
\multicolumn{2}{c}{M4.5e}\\
\( V \)         &12.16& 0.03&13.42& 0.02\\
\( B-V \)       & 1.50& 0.02& 1.58& 0.02\\
\( V-I_{c} \)   & 2.78& 0.02& 3.04& 0.01\\
\( J \)         & 7.78& 0.02& 8.68& 0.02\\
\( H \)         & 7.14& 0.04& 8.05& 0.04\\
\( K_{s} \)     & 6.91& 0.04& 7.77& 0.04\\
R.V. (km/sec)   &+2.5 & 0.6 &-1.7 & 2.4\\
EW(H\( \alpha  \)) \AA&
\multicolumn{2}{c}{\( -6.4 \)}&
\multicolumn{2}{c}{\( -4.3 \)}\\
EW(Li) m\AA&
\multicolumn{2}{c}{\( <30 \)}&
\multicolumn{2}{c}{\( 290 \)}\\
\( \mu _{\alpha } \) (mas/yr)&  179.8 & 5.2 &  179.8 & 2.9 \\
\( \mu _{\delta } \) (mas/yr)& -129.2 & 3.8 & -126.0 & 3.0 \\
TiO5 index&
\multicolumn{2}{c}{0.38}&
\multicolumn{2}{c}{0.34}\\
\hline
\end{tabular}\par}
\end{tablehere}
just a feature of young K and M dwarfs.  The \( V-I \)
and \( V-K \) colors are quite consistent and correspond to spectral
types around M4 and M4.5 for HIP~112312 A and B, respectively. 

We also measured the TiO5 spectral index (\( \equiv
F[7126-7135]/F[7042-7046] \)) for HIP~112312 A and B together with
that for GJ~644 (an M3 standard) following \citet{TiO5}. The
measured TiO5 index for GJ~644 (TiO5\( = \)0.48) was in good
agreement with its standard value and with its \( V-I \) color.
Based on the TiO5 and spectral type relation of \citeauthor{TiO5}
(\( Spectral\, type\, subclass=-10.775\times \mathrm{TiO}5+8.2 \)),
the TiO5 indices for HIP~112312 A \& B again indicate M4.1 and M4.5
spectral types, respectively. From colors and TiO5 indices, we
assign M4 and M4.5 spectral types to the primary and secondary,
respectively.

The effective temperatures of HIP~112312 A and B from their \( V-I \)
and \( V-K \) colors using an empirical temperature calibration and
model colors \citep{Bessell-Mstars} are estimated to be 3150~K and
3030~K respectively, and their uncertainties are not larger than 100~K.

\subsubsection{Lithium Abundance\label{LiContent}}
Estimation of Li abundances from observed Li~6708~\AA{} equivalent
width requires information on Li curves of growth. From synthetic
model spectra (\citealt{PHEONIX} and private communication with P.
Hauschildt for updates) for $T_{eff}$=3000, 3100, \& 3200~K and
$\log g$=4.0 \& 4.5 with a range of Li abundances ($\log N(Li)=-$99,
0.05, 0.1, 0.3, 0.5, 0.75, 1.0, \& 1.3), we constructed a set of
Li~6708\AA{} curves of growth (Figure~\ref{LiCurve}).  We note that
the indicated curves of growth take into account the effect of
strong TiO band absorption features around 6708\AA{} which can alter
the lithium equivalent width measurement depending on the location
of the pseudo-continuum level (Pavlenko 2002, priv. comm.). When our
Li curves of growth are compared with Pavlenko's (2002, priv. comm.)
model data which also takes the effect of TiO bands into account,
our curves of growth differ by $\lesssim15$\,m\AA{} at common Li
abundance points ($\log N(Li)=1.0\,\&\,1.3$). Using the estimated
effective temperature of the secondary from section \ref{Teff}
(3030~K) and the measured equivalent width of 290~m\AA{}, from
Figure~\ref{LiCurve} we conservatively estimate the Li abundance of
the secondary to lie in the range  $\log N(Li)=0.85-1.10$.   If we
adopt \( \log n(Li)=3.31 \) as the interstellar Li abundance, then
HIP~112312~A and B currently have only \( <0.01 \)~\% and \(
0.35-0.62 \)~\% of their initial Li contents.

\subsubsection{Luminosities}

It is quite straightforward to calculate luminosities of the
HIP~112312 binary system from Hipparcos measured parallax (\(
42.4\pm 3.4 \)~mas) and measured photometric magnitudes. In
calculation of bolometric magnitudes, we use \( m-M=1.86 \) for the
distance modulus, \( M_{bol}=4.75 \) for the Sun, bolometric
correction of 2.73 and 2.77 magnitude at K band for the primary and
the secondary, respectively \citep{BessellBC}.  The calculated
bolometric magnitudes are 7.78 \& 8.68 and the calculated
luminosities of A \& B components are \( \log (L/L_{\odot })=-1.22
\) and \( -1.58 \), respectively.  Uncertainty in luminosity is
estimated to be \( \Delta \log (L/L_{\odot })=0.160 \) which is
mainly due to the Hipparcos distance uncertainty.

\subsubsection{Space Motions \label{UVW}}

Physical companionship of the HIP~112312 components is verified from
the common proper motions of the primary and the secondary
(Table~\ref{data}).  The proper motions listed in Table~\ref{data}
are weighted means of SPM~2.0 \citep{SPM2} and UCAC1 \citep{UCAC1}
catalog values. Proper motions of the primary from Hipparcos catalog
(\( \mu _{\alpha }=183.12\pm 2.50 \)~mas/yr, \( \mu _{\delta
}=-118.87\pm 2.21 \)~mas/yr)  are in fair agreement with the Table~2
values.

From our echelle spectra, we determined weighted mean radial
velocities of the primary and secondary from a line fitting method
(see Table~\ref{data}).  We also estimated the rotational velocity (\(
v\sin i\)) of the primary to be (\( \sim 40 \)~km/sec) but were unable
to do so for the secondary because of its low S/N spectrum.  Using the
measured radial velocities, Hipparcos distances, and proper motions,
we derived galactic space motions (\( U,V,W \)) with respect to the
Sun of HIP~112312 by using the formulation of \citet{Johnson87}; \(
(U,V,W)=(-11.8\pm 1.2,-19.0\pm 1.6,-11.0\pm 0.9)\,km/sec \) for the
primary and $(-13.8\pm1.5, -19.1\pm1.6,-7.2\pm2.2)\,km/sec$ for the
secondary.  Based on the similar space motions, proximity to Earth
(24~pc), and very young ages (\( \lesssim 20 \)~Myr, see
Figure~\ref{HRD} and discussion in section \ref{ages}), the HIP~112312
binary components may be members of the \( \beta \)~Pictoris moving
group \citep{bPic2}.  The \( \beta \)~Pictoris group was defined as
nearby stars that have UVW velocity each within \( \sim 2 \)~km/sec of
those of \( \beta  \)~Pictoris, \( (-10.8,-16.4,-8.9) \), and age
consistent with 12~Myr \citep{bPic2}.

\subsection{Discussion and Comparison to Models\label{ages}}

We compare pre-main sequence (PMS) stellar evolutionary models with
our observed HIP~112312 binary data by plotting HIP~112312~A and B on
a \( \log T_{eff} \) versus \( \log (L/L_{\odot }) \) plot along with
three PMS models (Figure~\ref{HRD}).  The high luminosities of A and B
clearly indicate their young age (\( <20 \)~Myr); however the scatter
among the estimated ages from different models is rather large. 

The three theoretical models plotted in Figure~\ref{HRD} also predict
Li depletion patterns for PMS stars. Since we know how much Li
currently remains in HIP~112312 A and B components (section
\ref{LiContent}), we can directly compare the observed Li contents to
the predicted values (Table~\ref{LiTable}).

\citet{bPic2} identified \( \sim 30 \) co-moving stars near Earth, the
\( \beta \)~Pictoris moving group. By plotting late-type members on a
color magnitude diagram with theoretical isochrones, they estimated an
age of the group as \( \sim 12 \)~Myr. 

\begin{figurehere}
{\centering \resizebox*{0.9\columnwidth}{!}{\includegraphics{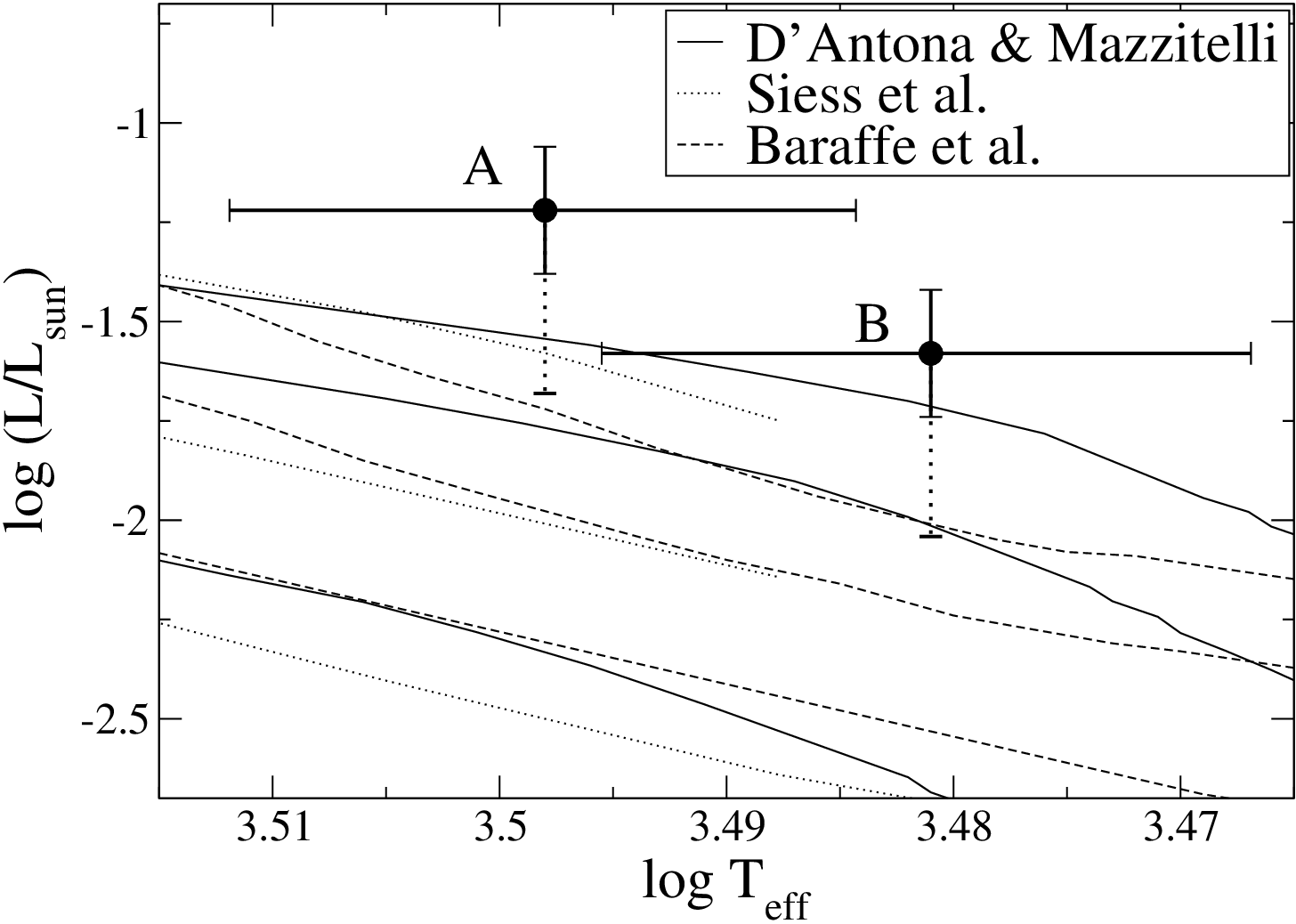}} \par}
\caption{Hertzsprung-Russell diagram of the A and B components of
HIP~112312 with isochrones from three different PMS theoretical
stellar evolutionary models \citep{BCAH,DM97,Siess}.\label{HRD} From
top to bottom, 10, 20, \& 100~Myr isochrones are plotted for each
model. The dotted luminosity error bars indicate the case if either
A or B should itself happen to be an equal mass binary. However, to
the Adaptive Optics resolution at Keck, neither the primary nor the
secondary is a binary (Macintosh et al., in prep.). If HIP~112312 A
\& B are $\sim12$~Myr old, then their masses are $\sim0.32\,M_\odot$
and $\sim0.18\,M_\odot$, respectively, based on the above models.}
\end{figurehere}

\begin{table*}
\caption{Predicted and observed Li contents for stars with $\log L/L_\odot=-1.22$ (primary) and $\log L/L_\odot=-1.58$ (secondary). \label{LiTable}}
{\centering \begin{tabular}{lcrr}
\hline 
\multicolumn{1}{c}{Model}& age&
\multicolumn{2}{c}{$^7$Li content (\% of initial content)}\\
& (Myr)& Primary& Secondary\\
\hline
\hline 
Observed values& \( <20 \)& \( <0.01 \)& \( 0.35-0.62 \)\\
\hline
\citet{BCAH} & 10& $\sim98$ & 100      \\
             & 20& $\sim0.7$& $\sim56$ \\
\hline
\citet{DM97} & 10& $\sim85$ & 100      \\
             & 20& $<0.01$  & $\sim5$  \\
\hline
\citet{Siess}& 10& $\sim98$ & 100      \\
             & 20& $<0.01$  & $\sim69$ \\
\hline
\end{tabular}\par}
\tablecomments{Among different theoretical models, \citet{DM97} fits
the observed data best and all LDB ages of open clusters were based
on the \citeauthor{DM97} model. However, even at the age upper limit
of the HIP 112312 binary system, the \citeauthor{DM97} model still
predicts too slow Li depletion ($\sim5$\,\% of Li content for the
secondary corresponding to $>600$\,m\AA{} of Li~6708\,\AA{}
equivalent width).}
\end{table*}

Recently, \citet{Brazil} estimated a kinematic age of the \( \beta  \)~Pictoris moving group by
tracing positions of individual members back in time. They found that
the \( \beta  \)~Pictoris members occupied the smallest volume about
11.5~Myr ago; hence they assigned a kinematic age of 11.5~Myr.  We
obtain somewhat similar results when we trace positions of the
currently known members back in time using constant velocity
trajectories (i.e., no galactic potential treatment, Song et al. in
prep.). Many members were more concentrated into a central cluster
about 12~Myr ago although the overall size of the moving group has not
been reduced as much as claimed in \citet{Brazil}.  The reasonably
good agreement between photometric isochrone age and dynamical age of
the \( \beta \)~Pictoris group indicates that ages estimated from
theoretical isochrones can be trusted in general for \( \sim 10 \)~Myr
old stars. 

Based on the facts that the HIP~112312~A component locates on the
Hertzsprung-Russell diagram above the 10~Myr isochrone and that
HIP~112312 AB may belong to a large \( \sim 12 \)~Myr old \( \beta
\)~Pictoris moving group \citep{bPic2,Brazil}, we believe that the
true age of HIP~112312 cannot be older than 20~Myr. At the age upper
limit of 20~Myr, all three theoretical models predict too slow Li
depletion (here we assume that for given ages, theoretical model
effective temperature and luminosity calculations are more reliable
than lithium depletion calculations). If the true age of
HIP~112312 is younger than $\sim12$~Myr (as may be hinted from
Figure~\ref{HRD}), then inconsistency between photometric data and Li
depletion timescales of contemporary theoretical models becomes more
severe. If the true age of HIP~112312 is not significantly different
from $\sim12$~Myr, thus requiring a faster Li depletion process than
those predicted by contemporary models to fit the observed data, then
ages of open clusters (Pleiades, \( \alpha \)~Persei, and IC 2391)
estimated from the LDB method are all overestimated. Since an age
scale of open clusters is critical and tied into many areas of
astronomy, developing self-consistent theoretical models to match the
HIP~112312 binary system data is an urgent task.

Lastly, it is worthwhile mentioning that a low mass secondary
component (M4.5) of a previously identified \( \beta  \)~Pictoris
member (V343~Nor, K0V) also shows a similarly strong Li~6708~\AA{}
feature (Song et al, in prep.). V343~Nor~B and HIP~112312~B have
almost the same spectral types based on the TiO5 spectral indices and
photometric colors. Study of V343~Nor~B corroborates the idea of
HIP~112312 being a member of the \( \beta  \)~Pictoris moving group
and the location of the LDB must be at \( \sim  \)M4.5 for \( 10-20
\)~Myr old stars.

\section{Summary and Discussion}

We have identified the Li depletion boundary in a \( <20 \)~Myr old
pre-main sequence binary system, HIP~112312 A \& B. From its galactic
space motion, young age, and proximity to Earth, the HIP~112312 binary
system may be a member of the \( \sim 12 \)~Myr old \( \beta
\)~Pictoris moving group. At the conservative age upper limit of 20~Myr,
all contemporary theoretical evolutionary models predict too slow Li
depletion.  Incorrect model Li depletion rates may cause overestimated
open cluster ages (Pleiades, \( \alpha  \)~Persei, and IC~2391) from the
lithium depletion boundary method. Because of the importance of the open
cluster age scale, development of self-consistent theoretical models to
satisfy HIP~112312 binary system data is desirable.

\acknowledgements{We thank Isabelle Baraffe, Eric Becklin, Andrea
Ghez, and Yakiv Pavlenko for helpful comments on the manuscript and
Peter Hauschildt for calculating synthetic spectra. We also
acknowledge referee John Stauffer for helpful comments and criticisms.
This research was supported by the UCLA Astrobiology Institute and by
a NASA grant to UCLA.}


\begin{thebibliography}{20}
\expandafter\ifx\csname natexlab\endcsname\relax\def\natexlab#1{#1}\fi

\bibitem[Allard et al.(2001)]{PHEONIX} Allard, F., Hauschildt,
P.~H., Alexander, D.~R., Tamanai, A., \& Schweitzer, A.\ 2001, \apj, 556,
357

\bibitem[\protect\astroncite{{Baraffe} {\em et~al.\/}}{1998}]{BCAH}
{Baraffe}, I., {Chabrier}, G., {Allard}, F., \& {Hauschildt}, P.~H. 1998, {\em
  \aap\/}, {\bf 337}, 403

\bibitem[\protect\astroncite{{Barrado y Navascu{\' e}s} {\em
  et~al.\/}}{1999}]{Barrado99}
{Barrado y Navascu{\' e}s}, D., {Stauffer}, J.~R., \& {Patten}, B.~M. 1999,
  {\em \apjl\/}, {\bf 522}, L53

\bibitem[\protect\astroncite{{Basri}}{2000}]{BasriARAA}
{Basri}, G. 2000, {\em \araa\/}, {\bf 38}, 485

\bibitem[\protect\astroncite{{Basri} {\em et~al.\/}}{1996}]{Basri}
{Basri}, G., {Marcy}, G.~W., \& {Graham}, J.~R. 1996, {\em \apj\/}, {\bf 458},
  600

\bibitem[Bessell(1991)]{Bessell-Mstars} Bessell, M.~S.\ 1991, \aj,
101, 662

\bibitem[\protect\astroncite{{Bessell} {\em et~al.\/}}{1998}]{BessellBC}
{Bessell}, M.~S., {Castelli}, F., \& {Plez}, B. 1998, {\em \aap\/}, {\bf 333},
  231

\bibitem[\protect\astroncite{{Bildsten} {\em et~al.\/}}{1997}]{Bildsten}
{Bildsten}, L., {Brown}, E.~F., {Matzner}, C.~D., \& {Ushomirsky}, G. 1997,
  {\em \apj\/}, {\bf 482}, 442

\bibitem[\protect\astroncite{{D'Antona} \& {Mazzitelli}}{1997}]{DM97}
{D'Antona}, R.~A. \& {Mazzitelli}, I. 1997, in {\em Cool stars in Clusters and
  Associations\/}, edited by R.~{Pallavicini} \& G.~{Micela}, vol.~68 of {\em
  Mem.S.A.It.\/},  807

\bibitem[\protect\astroncite{{Johnson} \& {Soderblom}}{1987}]{Johnson87}
{Johnson}, D.~R.~H. \& {Soderblom}, D.~R. 1987, {\em \aj\/}, {\bf 93}, 864

\bibitem[\protect\astroncite{{Kenyon} \& {Hartmann}}{1995}]{KH95}
{Kenyon}, S.~J. \& {Hartmann}, L. 1995, {\em \apjs\/}, {\bf 101}, 117

\bibitem[\protect\astroncite{{Ortega} {\em et~al.\/}}{2002}]{Brazil}
{Ortega}, V.~G., {de la Reza}, R., {Jilinski}, E., \& {Bazzanella}, B. 2002,
  {\em \apjl\/}, {\bf 575}, 75

\bibitem[\protect\astroncite{{Platais} {\em et~al.\/}}{1998}]{SPM2}
{Platais}, I., {\em et~al.\/} 1998, {\em \aj\/}, {\bf 116}, 2556

\bibitem[\protect\astroncite{{Reid} {\em et~al.\/}}{1995}]{TiO5}
{Reid}, I.~N., {Hawley}, S.~L., \& {Gizis}, J.~E. 1995, {\em \aj\/}, {\bf 110},
  1838

\bibitem[\protect\astroncite{{Siess} {\em et~al.\/}}{2000}]{Siess}
{Siess}, L., {Dufour}, E., \& {Forestini}, M. 2000, {\em A\&A\/}, {\bf 358},
  593

\bibitem[\protect\astroncite{{Stauffer} {\em et~al.\/}}{1998}]{Stauffer98}
{Stauffer}, J.~R., {Schultz}, G., \& {Kirkpatrick}, J.~D. 1998, {\em \apjl\/},
  {\bf 499}, L199

\bibitem[\protect\astroncite{{Stauffer} {\em et~al.\/}}{1999}]{Stauffer99}
{Stauffer}, J.~R., {\em et~al.\/} 1999, {\em \apj\/}, {\bf 527}, 219

\bibitem[\protect\astroncite{{Ventura} {\em et~al.\/}}{1998}]{Ventura}
{Ventura}, P., {Zeppieri}, A., {Mazzitelli}, I., \& {D'Antona}, F. 1998, {\em
  \aap\/}, {\bf 331}, 1011

\bibitem[\protect\astroncite{{Zacharias} {\em et~al.\/}}{2000}]{UCAC1}
{Zacharias}, N., {\em et~al.\/} 2000, {\em \aj\/}, {\bf 120}, 2131

\bibitem[\protect\astroncite{{Zuckerman} {\em et~al.\/}}{2001}]{bPic2}
{Zuckerman}, B., {Song}, I., {Bessell}, M.~S., \& {Webb}, R.~A. 2001, {\em
  ApJL\/}, {\bf 562}, L87

\end{thebibliography}
\end{document}